\documentclass[sn-nature,Numbered]{sn-jnl}

\usepackage{graphicx}%
\usepackage{multirow}%
\usepackage{amsmath,amssymb,amsfonts}%
\usepackage{amsthm}%
\usepackage{mathrsfs}%
\usepackage[title]{appendix}%
\usepackage{xcolor}%
\usepackage{textcomp}%
\usepackage{manyfoot}%
\usepackage{booktabs}%
\usepackage{algorithm}%
\usepackage{algorithmicx}%
\usepackage{algpseudocode}%
\usepackage{listings}%

\usepackage{multibib}
\newcites{supp}{Methods References}
\usepackage{caption}

\theoremstyle{thmstyleone}%
\theoremstyle{thmstyletwo}%

\theoremstyle{thmstylethree}%

\usepackage{aas_macros}

\def\deg{\hbox{$^\circ$}}

\def\lesssim{\mathrel{\hbox{\rlap{\hbox{\lower4pt\hbox{$\sim$}}}\hbox{$<$}}}}
\def\gtrsim{\mathrel{\hbox{\rlap{\hbox{\lower4pt\hbox{$\sim$}}}\hbox{$>$}}}}

\begin{document}

\title[]{One month convection timescale on the surface of a giant evolved star}

\author*[1]{\fnm{Wouter} \sur{Vlemmings}}\email{wouter.vlemmings@chalmers.se}

\author[1]{\fnm{Theo} \sur{Khouri}}\email{theo.khouri@chalmers.se}

\author[1]{\fnm{Behzad} \sur{Bojnordi Arbab}}\email{bojnordi@chalmers.se}
\author[1]{\fnm{Elvire} \sur{De Beck}}\email{elvire.debeck@chalmers.se}
\author[1]{\fnm{Matthias} \sur{Maercker}}\email{matthias.maercker@chalmers.se}

\affil*[1]{\orgdiv{Department of Space, Earth and Environment}, \orgname{Chalmers University of Technology}, \orgaddress{\postcode{412 96}, \city{Gothenburg},  \country{Sweden}}}

\maketitle

\textbf{\footnote{This version of the article has been accepted for publication, after peer review but is not the Version of Record and does not reflect post-acceptance improvements, or any corrections. The Version of Record is available online at: http://dx.doi.org/10.1038/s41586-024-07836-9} The transport of energy through convection is important during
  many stages of stellar evolution \cite{2023Galax..11...89L,
    2017LRCA....3....1K}, and is best studied in our Sun
  \cite{2009LRSP....6....2N} or giant evolved stars
  \cite{1975ApJ...195..137S}. Features that are attributed to
  convection are found on the surface of massive red supergiant stars
  \cite{2011ApJ...740...24R, 2017A&A...605A.108M, 2020A&A...635A.160C,
    2021ApJ...919..124N}. Also for lower mass evolved stars,
  indications of convection are found \cite{2016A&A...587A..12W,
    2018Natur.553..310P, 2019A&A...626A..81V, 2024arXiv240213676K,
    2023Natur.617..696V}, but convective timescales and sizes remain
  poorly constrained. Models indicate that convective motions are
  crucial for the production of strong winds that return the products
  of stellar nucleosynthesis into the interstellar medium
  \cite{2023A&A...669A.155F}.  Here we report a series of
  reconstructed interferometric images of the surface of the evolved
  giant star R~Doradus. The images reveal a stellar disc with
  prominent small scale features that provide the structure and
  motions of convection on the stellar surface. We find that the
  dominant structure size of the features on the stellar disc is
  $0.72\pm0.05$ astronomical units (au). We measure the velocity of
  the surface motions to vary between $-18$ and $+20~$km~s$^{-1}$,
  which means the convective timescale is approximately one
  month. This indicates a possible difference between the convection
  properties of low-mass and high-mass evolved stars.}

\bigskip


The M-type asymptotic giant branch (AGB) star R~Doradus (see also the
Methods section~\ref{source}) was observed with the Atacama Large
Millimeter/submillimeter Array (ALMA) in five epochs spread over four
weeks between 2023 July 2 and August 2. The observations around
338~GHz were done using the longest available ALMA baselines. Using
{\it superuniform} weighthing, this allowed us to reconstruct images
of the stellar surface at an angular resolution between $8-25$~mas
(see Method section~\ref{obs}). At submillimeter wavelengths, the main
opacity source are the electron-neutral free-free interactions
\cite{1997ApJ...476..327R}. As a result, the observations are
sensitive to the temperature and density structure in an extended
atmosphere above the stellar photosphere. In particular, submillimeter
observations probe the dynamics and properties of the shocks excited
by large convective cells on the stellar photosphere, while not being
hindered by dust or molecular opacity sources that dominate at other
wavelengths \cite{2019A&A...626A..81V}. The surface maps for the three
observational epochs with the highest angular resolution are shown in
Fig.~\ref{fig: epochs}, the remaining epochs are shown in the
supplementary material. The maps show a predominantly circular stellar
disc with a radius of $R_{\rm 338~GHz}=1.64\pm0.09$~au and a
brightness temperature of $T_b=2270\pm130$~K (see Methods
section~\ref{source}). Considering the correspondence between several
epochs, including observations taken at $225$~GHz shown in the Methods
section \ref{band6}, we conclude that the structures that we observe
are intrinsic to the star and are very likely induced by surface
convection. From the visual inspection, we find that the structures we
observe at the surface have a typical lifetime of at least three
weeks.

\begin{figure}[t!]
\centering
\includegraphics[width=0.8\textwidth]{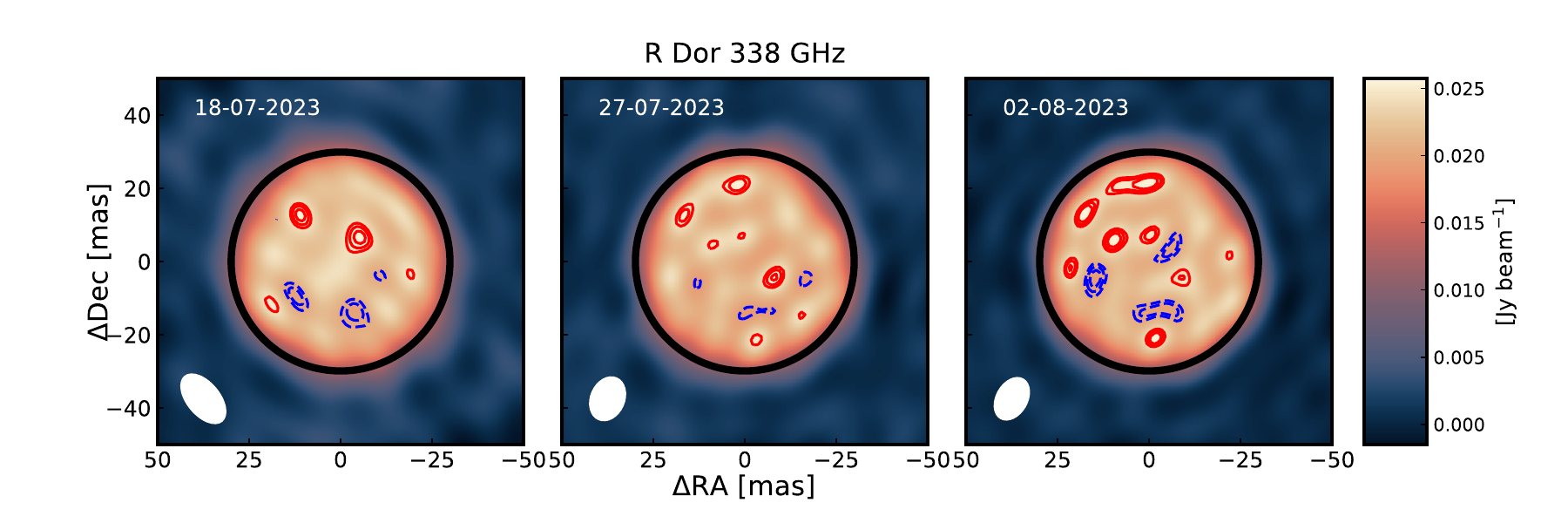}
\caption{The stellar surface of the AGB star R~Doradus. The
panels represent the three highest angular resolutions epochs of ALMA
observations at 338~GHz. The black ellipse in the panels indicates the
average size of the stellar disc at this frequency. The red solid
contours and the blue dashed contours indicate the positive and
negative $4, 5,$ and $6\sigma$ features with respect to the mean
emission of the stellar disc. The size and orientation of the
interferometric beam is indicated at the bottom left of each panel.}\label{fig: epochs}
\end{figure}

\begin{figure}[h!]
\centering
\includegraphics[width=0.8\textwidth]{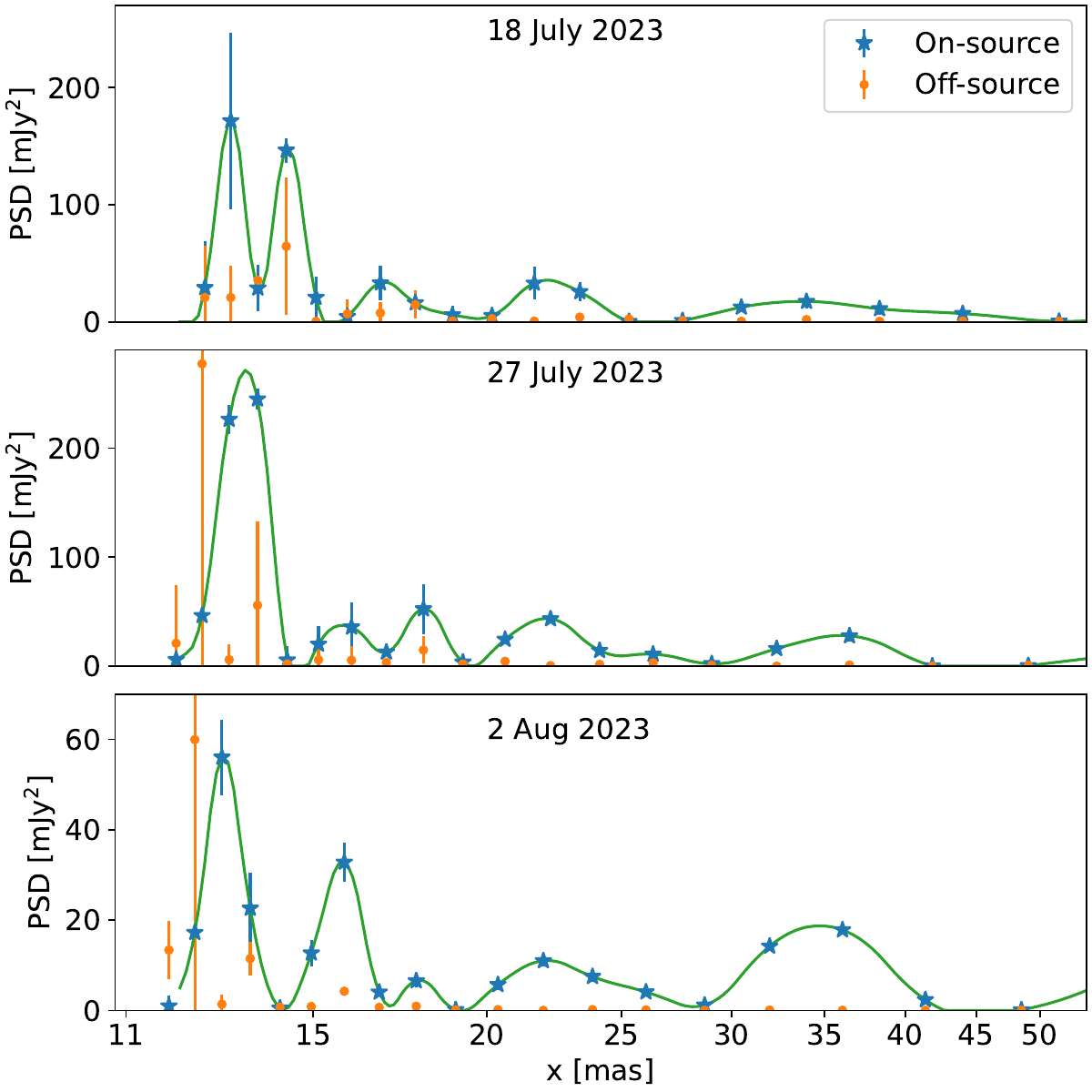}
\caption{The spatial power spectrum density for three epochs of
observations of R~Doradus. The spatial power spectrum density (PSD),
in units of mJy$^2$, was determined directly from the interferometric
visibilities after subtracting the best fit model for the stellar
disc. The three panels represent the final three, highest angular
resolution, epochs. The solid circles denote the measurements and the
curve represents a cubic spline fit to the observations. The largest blue
stars, with corresponding $1\sigma$ s.d. error bars, are the PSD
determined on-source, while the smaller orange dots represent the
off-source measurements. In the last two epochs, the second bin at
$\sim12$~mas contains only limited visibilities and hence does not
present a significant detection. As the peak at the smallest angular
size dominates, this indicates that the smallest granule structure
dominates. From an error weighted average of the three epochs, we find
a value of $x=13.1\pm0.6$~mas. For a distance of $55\pm3$~pc
\cite{2007A&A...474..653V} this corresponds to $x=0.72\pm0.05$~au,
compared to a stellar diameter at $338$~GHz of $3.28$~au.}\label{fig: psd}
\end{figure}

In order to estimate the size of the dominant structures on the
submillimeter surface, we employ a spatial power spectral density
(PSD) analysis (see the description in the Method
section~\ref{psd}). The PSD analysis was performed using the
calibrated interferometric visibilities after subtracting the best fit
model for the stellar disc in the three last, highest angular
resolution, epochs. We show the results in Fig.~\ref{fig: psd}. We
find that structure exists at several scales on the surface. The
structure varies between the three epochs, particularly in amplitude,
and most of the power is concentrated in small scale structure with a
size of $x=13.1\pm0.6$~mas. At a distance of R~Doradus of $55\pm3$~pc
\cite{2007A&A...474..653V}, this corresponds to a size of
$x=0.72\pm0.05$~au. Other peaks present in the power spectrum are
located at $16.0\pm1.0, 17.9\pm0.6, 22.1\pm1.3,$ and
$35.2\pm3.6$~mas. The largest of these correspond to more than half
the visible hemisphere. The different scales might be related to the
equivalent scales of granulation, mesogranulation and, considering the
scales, in particular supergranulation inferred for our Sun
\cite{2003ApJ...597.1200R, 2018LRSP...15....6R}, or could be the
result of a superposition of several independent granules. In the absence of a
better name we subsequently use the term granules for the smallest
structures, although concluding a direct link with Solar granules is
not yet possible. On average, the power on each of the four larger
scales is $\sim25\%$ of that in the dominant structure.

\begin{figure}[h!]
\centering
\includegraphics[width=0.8\textwidth]{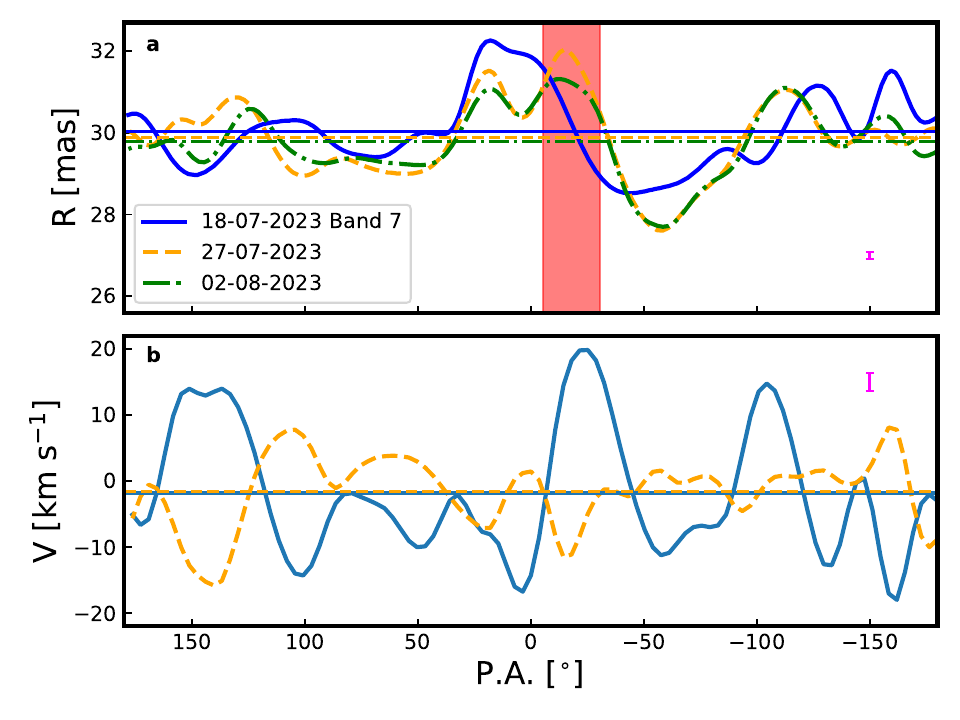}
\caption{The radius and radial velocity of the stellar surface
of R~Doradus. {\bf Panel a} The half-power radius $R$ as a function of
the position angle (P.A.) with respect to the north celestial
pole. Positive angles point in the direction of the right ascension.
The curves indicate $R$ measured for three epochs of ALMA Band 7
(338~GHz) observations. The horizontal lines indicate the average
radius for each epoch. The lower figure boundary on the radius is the
estimated photospheric radius of $25.6$~mas, measured at $2.3\mu$m
\cite{2019ApJ...883...89O}. The red vertical bar indicates the granule
size measured from the spatial power spectral density, with the linear
size translated to an angular size at $R$. The $1\sigma$ s.d. on the
radius (similar for each epoch) is plotted in the bottom right corner
(magenta errorbar) and is at most $0.17$~mas. {\bf Panel b} The radial
velocity $V$ at the surface of R~Doradus. The velocity is determined
from the difference of radius between the third and fourth epoch (18
and 27 July 2023; solid line) and fourth and fifth epoch (27 July and
2 Aug 2023; dashed line). The horizontal lines denote the average
velocity. The velocity determined in this way corresponds to the
movement of the $\tau_{338~GHz}=1$ optical depth surface and is an
average between the respective epochs. This indicates the movement of
the shocks induced by the convective motions. The $1\sigma$ s.d. in
the radius determination translates in an uncertainty on the velocity
of $\lesssim 2.6$~km~s$^{-1}$ and is indicated in the top right of the
panel (magenta errorbar).}\label{fig: radvel}
\end{figure}

The structures inside the stellar disc can also be compared with
structures seen on the limb of the star. In Fig.~\ref{fig: radvel}a,
we show the half-power radius as a function of position angle for the
three highest angular resolution epochs. There is a good
correspondence between the epochs, with specifically the last two
epochs showing very similar structures in the half-power radius in
Fig.~\ref{fig: radvel}a. As shown in the supplementary
information~\ref{band6}, also the 225~GHz observations show similar
features confirming that they are intrinsic to the star. By
determining the difference in radius between consecutive epochs at
every position angle, we can derive the average velocity profile of
the 338~GHz optical depth surface as shown in Fig.~\ref{fig:
  radvel}. Negative velocities indicate a radial motion inward to the
star. While the average velocity is small, the movement of the optical
depth surface as a function of angle varies between $V = -18$ and
$+20$~km~s$^{-1}$. These velocities can be compared with the local sound
speed $V_s\sim$6~km~s$^{-1}$ in the part of the extended atmosphere
probed by the observations and are consistent with supersonic shocks
induced by convection \cite[e.g.][]{2013A&A...557A...7T}.  Although
the observed motions depend on a combination of changes in gas
density, temperature, ionisation and velocity, they are, on short
timescales, a good representation of the (radial) motions of the
shocks. They thus represent a lower limit to the actual gas
velocities. The escape velocity at the
measured submillimeter radius for a star with the parameters adopted
for R~Doradus is $V_{\rm esc}=30\pm3$~km~s$^{-1}$. Hence, as a
consequence of these shocks, only a very small fraction of the gas
will be able to escape the gravity field of the star before being
accelerated by radiation pressure on dust \cite{2023A&A...669A.155F}.

\begin{figure}[h!]
\centering
\includegraphics[width=0.8\textwidth]{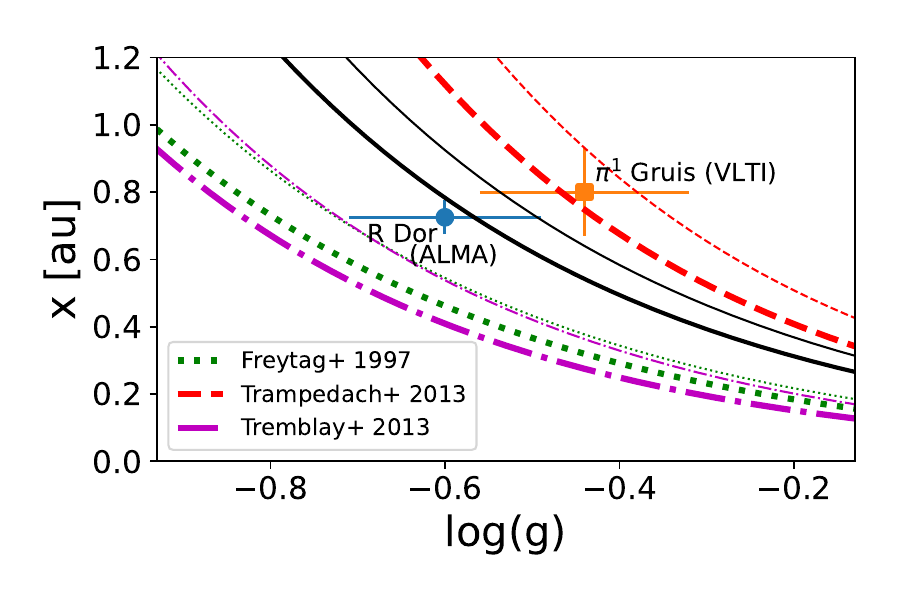}
\caption{The smallest size of the granules on the surface of
R~Doradus determined using ALMA compared with theoretical predictions
and the previous VLTI measurement for $\pi^1$~Gruis
\cite{2018Natur.553..310P}. The observations are plotted vs. $\log
(g)$ and the $1\sigma$ s.d. error bars include the uncertainty on the
distance and stellar mass. The solid black lines indicate the
parametric scaling presented in this paper. The thick line was
calculated for a stellar effective temperature $T_{\rm eff}=2710$~K,
which is the value for R~Doradus. The thin line is the same model for
$T_{\rm eff}=3200$~K, which is closer to the value for
$\pi^1$~Gruis. The other lines (dotted \cite{1997svlt.work..316F},
dashed \cite{2013ApJ...769...18T}, and dash-dotted
\cite{2013A&A...557A...7T}) represent parametric models from the
literature, with the thick and thin lines calculated for the stellar
temperatures of $2710$ and $3200$~K, respectively. For the model from
\cite{2013A&A...557A...7T} we assume solar metallicity.}\label{fig: xau}
\end{figure}

Considering the measurement of the size of the granules as
well as the velocity of the shocks, we can determine the timescale for
the surface structures to readjust
\cite[e.g.][]{2024ApJ...962L..36M}. This timescale is given by $t_{\rm
  surf}\sim x/\Delta V = 33\pm3$~days, which is independent of the
assumed distance to the star. This represents the first measurement of
this timescale, which corresponds well with the timescale of more than
three weeks estimated from the visual inspection of the images and is
consistent with model predictions \cite{2017A&A...600A.137F} but is
only $\sim30\%$ of the convective decay time extrapolated using the
emperical formula (Equation 16) from \cite{2013A&A...557A...7T}. As this emperical
formula was derived for models of stars with both higher effective
temperatures and surface gravity, our result points to a possible
difference of convection properties for AGB stars.

The contrast between the average brightness of the stellar disk and
the granules induced by the surface convective motions
varies between $2.8, 3.2,$ and $1.5\%$ during the last three epochs
(Fig.~\ref{fig: psd}). This is notably less than the $\sim12\%$
contrast observed for the star $\pi^1$~Gruis in the near-infrared. As
the contrast relates to the imprint of the convective motions on the
shock structure in the extended atmopshere, we can use the contrast
and size to determine the average brightness temperature increase
$\Delta T_b$ caused by these shocks. We find $\Delta T_b$ to be
between $\sim700$ and $1500$~K. As the brightness temperature is
closely related to the real temperature of the extended atmosphere
\cite[e.g.][]{2019A&A...626A..81V, 1997ApJ...476..327R}, this is
sufficient to affect the chemistry in the stellar atmosphere \cite[e.g.][]{2006A&A...456.1001C}. Previous
observations have also revealed brighter and more compact hotspots on
the sub-millimeter surface of AGB stars, including R~Dor
\cite[e.g.][]{2015A&A...577L...4V, 2017NatAs...1..848V,
  2019A&A...626A..81V}. These hotspots had a contrast in excess of
$5\%$ which, because of their compact size ($<0.3$~au), corresponded
to an increase in brightness temperature that can reach $\Delta
T_b > 50000$~K. These hotspots have different characteristics from the
convective shock structures described here, and are more rare. Further
monitoring observations are needed to determine the timescales related
to the bright hotspots and characterise their origin.

Although in the future, our observations can serve as a benchmark
for the models of convection in M-type AGB stars, it is currently only
possible to compare the results with parametric formulas that predict
the granular size scale based on fundamental stellar properties such
as effective temperature, surface gravity and chemical
composition. All relations are based on mixing length theory and
extrapolated models of less evolved stars, and there are no models yet
that specifically aim to reproduce R~Doradus. In Fig.~\ref{fig: xau},
we compare our measurements (using the stellar parameters described in
the Method section~\ref{source}) with the three relations that were
also compared to the measurements of $\pi^1$~Gruis
\cite{2018Natur.553..310P}. While there is generally good agreement,
none of the relations is a perfect fit.  The size of the granulation
is generally thought to scale with the pressure scale height $H_p$
immediately below the photosphere \cite{1975ApJ...195..137S}, with
$x=\alpha H_p$. For AGB stars, the scaling parameter $\alpha$ is
assumed to be $\sim10$ \cite{1997svlt.work..316F}. For the more
massive red supergiant stars, $\alpha\sim30-50$ appears to better
match the hydrodynamical simulations and observations
\cite{2009A&A...506.1351C, 2021ApJ...919..124N}. Using $H_p = k_{\rm
  B}T/\mu g$, where $k_{\rm B}$ is the Boltzmann constant, $\mu$ is
the mean molecular mass and $g$ is the acceleration due to surface
gravity, this leads to the relation:
\begin{equation}
  x=0.427\left({{\alpha}\over{10}}\right)~\left({{T_{\rm eff}}\over{2500~{\rm K}}}\right)~\left({{g}\over{0.25~{\rm cm~s}^{-2}}}\right)^{-1}~[{\rm au}]
\end{equation}
Using the two AGB measurements, we can estimate a value of $\alpha=17^{+5}_{-3}$ that represents the best fit to both observations.

Our results indicate that the convective structure on the surface of
AGB stars matches the existing parametric formulas, in size but reveal
a difference between AGB stars and the more massive red
supergiants. The results match the timescales found in AGB models, but
the timescales differ from extrapolations of models based on less
evolved stars and thus serve as a unique benchmark for the existing
theory. The derived convection properties can also be implemented in
stellar evolution and population synthesis models where convection is
generally captured in a free parameter that has a large effect on
galaxy evolution models \cite[e.g.][]{2024arXiv240307414L}.

\newpage

\backmatter

\bmhead{Acknowledgements} WV, TK, and BB acknowledge support from the
Olle Engkvist foundation through grant nr. 229-0368 and from the
Swedish Research Council under grants No. 2014-05713 and 2019-03777.
We also acknowledge support from the Nordic ALMA Regional Centre
(ARC) node based at Onsala Space Observatory. The Nordic ARC node is
funded through Swedish Research Council grant No 2019-00208.
This paper makes use of the following ALMA data: ADS/JAO.ALMA\#2022.1.01071.S. ALMA is a partnership of ESO (representing its member states), NSF (USA) and NINS (Japan), together with NRC (Canada), NSC and ASIAA (Taiwan), and KASI (Republic of Korea), in cooperation with the Republic of Chile. The Joint ALMA Observatory is operated by ESO, AUI/NRAO and NAOJ.

\bmhead{Author contributions}
WV wrote the paper and performed the analysis of the data. TK wrote
the observing proposal.  WV, TK, BB, EDB, and MM contributed to the
interpretation of the results. The authors declare that they have no
competing financial or non-financial interests. Correspondence and
requests for materials should be addressed to Wouter Vlemmings
(\href{mailto:wouter.vlemmings@chalmers.se}{wouter.vlemmings@chalmers.se}). \\

\bmhead{Data availability}

The ALMA data are publicly available on the ALMA archive (\url{https://almascience.eso.org/aq/}) as part of project 2022.1.01071.S. The spatial power spectral density (PSD) for the three epochs is available in CSV file {\it 'RDor-psd.csv'}. This constitutes the source data for Fig.~\ref{fig: psd}. The half-power radius and velocity that is the source data for Fig.~\ref{fig: radvel} is available in the CSV file {\it 'RDor-radiusvelocity.csv'}. The radial profile of R~Doradus from Extended Data Fig.~\ref{fig: disc} is provided in the CSV file {\it 'RDor-profile.csv'}.

\bmhead{Code availability}

This paper makes use of the NRAO CASA package and python. The code used for $uv$-fitting {\it 'uvmultifit'} is publicly available.


\clearpage

\captionsetup[table]{name=Extended Data Table}
\captionsetup[figure]{name=Extended Data Figure}

\section*{Methods}\label{methods}

\subsection*{Source properties}
\label{source}

R~Doradus is an M-type AGB star that belongs to the class of
semi-regular pulsators. On a time-scale of $\sim1000$~days it switches
between two pulsation modes with periods of $362$ and $175$~days. The
distance to R~Doradus has been determined to be $55\pm3$~pc using revised {\it
  Hipparcos} measurements
\cite{2007A&A...474..653V}. There is no useable {\it Gaia}
parallax. From CO observations, it was determined that R~Doradus has a
relatively low mass loss rate ($\sim10^{-7}$~M$_\odot$~yr$^{-1}$) and
wind expansion velocity ($\sim5.7$~km~s$^{-1}$)
\citesupp{2014A&A...566A.145R}). Previous ALMA observations also indicated
that R~Doradus rotates fast for a giant star, with a rotation velocity
at the surface of $\sim1.0\pm0.1$~km~s$^{-1}$ compared to a rotation
velocity of a few 10s of m~s$^{-1}$ expected for solitary AGB stars
\cite{2018A&A...613L...4V}. It has been suggested that the apparent
rotation could be the result of a chance alignment of convective cells
\cite{2024ApJ...962L..36M}. However, the rotation has been observed in
multiple molecular lines at at least four different epochs which,
including the observations presented here, span more than six years
\cite[e.g.][]{2018A&A...613L...4V, 2024arXiv240213676K}. This is much
longer than the convective timescales found in our analysis, and hence
a chance alignment of convective cells can be ruled out.

In our comparison with the convective theory, we adopt the values for
the effective temperature of $T_{\rm eff}=2710\pm70$~K. For the surface
gravity we use $\log(g)=-0.6\pm0.1$, based on models that indicate the
initial mass was $1-1.25~M_\odot$ and that the current mass is
$0.7-1.0~M_\odot$, combined with interferometric measurements that
yield a stellar diameter in the infrared of $D_{\rm
  IR}=51.18\pm2.24$~mas \cite{2019ApJ...883...89O}. It is expected
that this diameter, which corresponds to a radius of $R_{\rm
  IR}=1.4\pm0.1$~au~$=298\pm21~R_\odot$, indicates the size of the
stellar photosphere. We can compare this with the ($\tau=1$) size of the star
determined with ALMA at $338$~GHz obtained by visibility
fitting. Using a combination of the last three epochs, we fit a nearly
circular stellar disc with $F_{\rm 338GHz}=521\pm18$~mJy, $D_{\rm
  338GHz}=59.8\pm0.4$~mas and an axis ratio of $0.99\pm0.01$. This
means a brightness temperature $T_b=2270\pm130$~K and, taking into
account the uncertainty on the distance, a radius $R_{\rm 338GHz} =
1.64\pm0.09$~au~$=353\pm19~R_\odot = 1.18\pm0.11~R_{\rm IR}$.

\subsection*{Observations, data reduction and imaging}
\label{obs}

The AGB star R~Doradus was observed in ALMA Bands 6 and 7 as part of
ALMA project 2022.1.01071.S (PI: Khouri). The Band 7 observations were
taken between July 5 and August 2 2023 using four spectral windows
(spws) centered at $331.2, 333.0, 342.1,$~and $345.1$~GHz. Each spw
had a bandwidth of $1.875$~GHz and 1920 channels. The integration time
of the individual visibilities was set to $2.02$~s. The observations
were taken in the largest ALMA configurations (C-9 and C-10) with the
quasars J0519-4546 and J0516-6207 as bandpass/amplitude and phase
calibrator, respectively. Details of the observations are presented in
Table~\ref{tab: Obs}.  The calibration of the last three epochs was
performed using the ALMA pipeline in CASA v6.4.1.12
\citesupp{2023PASP..135g4501H}. The first two epochs were labelled as {\it
  semi-pass} in the ALMA quality assurance and for these the
calibration was performed manually by staff from the Nordic ALMA
regional center node using CASA v6.5.4.9. In the first epoch, there
was an issue with the bandpass calibration that needed to be solved
manually. In the second epoch, one of the antennas needed to be
flagged, resulting in a loss of some of the longest baselines. For
both epochs, the requested angular resolution and sensitivity were not
reached. After the initial calibration of each epoch, molecular lines
were identified and flagged before the data was averaged to an
integration time of 6.06~s and to 50 channels per spw. Subsequently,
two steps of phase-only self-calibration was performed on the stellar
continuum. The self-calibration improved the signal-to-noise ratio
(snr) by a factor of $\sim2.5$ on the continuum. Finally, images,
using all four spw, were produced for the five epochs using {\it
  superuniform} visibility weighting \citesupp{2022PASP..134k4501C}. This
method increases the relative weight of the visibilities at the longer
baselines, which minimises the beamsize at the expense of snr. The
{\it superuniform} beam characteristics and continuum rms noise are
also given in Table~\ref{tab: Obs}. The increase of rms noise compared
to more regular {\it Briggs} weighting (with a robust parameter of
0.5) depends on the telescope distribution and was a factor of $\sim
2, 1.5, 1.5, 3,$ and $7$ for the five epochs, respectively. The
improvement in angular resolution between {\it superuniform} and {\it
  uniform} weighting ranged from $\sim 2\%$ in the final epoch to
$\sim 15\%$ in the third epoch. The {\it superuniform} weighted images
of the final three epochs are presented in Fig.~\ref{fig: epochs} and
the first two epochs in Extended Data Fig.~\ref{fig: firstepochs}. The images of the
highest angular resolution epochs were used to derive the angular
radial profile presented in Fig.~\ref{fig: radvel}. We verified that
reducing the angular resolution to match that of the epoch with the
largest beam smooths out the observed structures and the derived
velocities and thus use the highest angular resolution results.  The fits of the
stellar disc and the spatial power spectral density analysis were
performed on the calibrated visibilities. While we focus our analysis
on the higher resolution Band 7 observations, we also include the
continuum result from the Band 6 observations in the Methods section~\ref{band6}. The observational details for these observations
are also included in Table~\ref{tab: Obs} and the calibration,
self-calibration, and imaging steps were identical to those performed
for the Band 7 observations. The four spws are centered at $218.9,
220.8, 230.0,$~and $232.9$~GHz and the increase in the continuum rms
noise between {\it Briggs} weighting and {\it superuniform} weighting
is a factor $1.5$.

\subsection*{Spatial Power Spectral Density analysis}
\label{psd}

The spatial power spectral density (PSD) is regularly used to derive
information about e.g. the turbulent structure of the interstellar
medium \citesupp[e.g.][]{1983A&A...122..282C, 1993MNRAS.262..327G} as well
as the convective structure of the solar photosphere
\citesupp[e.g.][]{2009A&A...503..225W}. The spatial PSD is given by the 2D
Fourier transform of an image. But considering interferometric images
are themselves the Fourier transform of the interferometric
visibilities, the spatial PSD is equal to the modulus squared of the
complex visibilities \citesupp{1983A&A...122..282C}. We can thus calculate
the PSD directly from our interferometric visibilities without
introducing potential artifacts during the imaging process. Because
the PSD would be dominated by the power at the scales of the stellar
disc, we first use the $uv-$fitting code {\it uvmultifit}
\citesupp{2014A&A...563A.136M} to fit the stellar disc (for a discussion
on the disc profile see Section~\ref{disc}). Subsequently, we subtract
the disc from the visibilities after which we calculate, for a phase
center towards the center of the star, the PSD using visibilities
annularly averaged in equally spaced bins of $uv-$distance (in units
of k$\lambda$). From the inverse of the $uv-$distance we can directly
obtain the angular scale $x$ in milliarcseconds. In addition, we also
determine the PSD for a position offset from the star. We chose an
offset of $7$~arcseconds to avoid a contribution to the off-source PSD
from the source signal. Since the first two epochs have worse spatial
resolution, we only produce the PSD for the final three epochs. The
results are shown in Fig.~\ref{fig: psd}. By comparing the on- and
off-source PSD we can determine which structures are significant and
which are possibly due to correlated noise in the visibilities.

\subsection*{Stellar disc profile}
\label{disc}

In order to check how well a top-hat shaped stellar disc profile fits
the observations we have also investigated the stellar disc profile in
the image plane. We produced a radially averaged profile of R~Doradus
based on the combined data for the final three epochs. We then
compared this profile with a top-hat shaped stellar disc model
convolved with the interferometric beam. The results of this
comparison are shown in Extended Data Fig.~\ref{fig: disc}. The stellar disc model
can accurately describe the observations with residuals at a level of
$\sim 2\%$ of the peak emission. This means that the 338~GHz opacity
$\tau_{\rm 338 GHz}$ increases steeply over only a small change in
radius. The radial motions we observe thus reflect the physical motion
of the 338~GHz optical depth surface. Since the optical depth is a
strong function of the density \cite{2019A&A...626A..81V}, the motions
closely reflect the motion of the shocks induced by the
convection.

\subsection*{Band 6 observations}
\label{band6}

The ALMA Band 6 (225~GHz) observations and the radius as a function of
position angle are shown in Extended Data Fig.~\ref{fig: band6} and Extended Data \ref{fig:
  band6R} respectively. Fitting the interferometric visibilities yields
a completely circular stellar disc with $F_{\rm
  225GHz}=221.4\pm0.1$~mJy and $D_{\rm 225GHz}=61.8\pm0.1$~mas. This
means a brightness temperature $T_b=2006\pm8$~K and, including the
distance uncertainty, a radius $R_{\rm 225GHz} =
1.70\pm0.09$~au~$=365\pm20~R_\odot = 1.22\pm0.11~R_{\rm IR}$. In a
comparison between the radii of the 225~GHz observations and the last
epochs of the 338~GHz observations in Fig.~\ref{fig: band6R}, it is
clear that there is a very good correspondence. Since the different
observations are completely independent, this shows that the pattern
seen in the radii is intrinsic to the source at the time of the
observations.

\setcounter{figure}{0}    

\clearpage

\begin{table}[h]
\caption{Observational details}\label{tab: Obs}%
\begin{tabular}{@{}lllll@{}}
\toprule
Obs. date & Total/on-source time  & Min/max baseline length  & Beam size\footnotemark[1] & rms noise\footnotemark[1] \\
 & [hh:mm / hh:mm] & [m / m] & [mas$\times$mas, $\deg$] & [mJy~beam$^{-1}$] \\
\midrule
05-07-2023\footnotemark[2]    & 1:41/0:36  & 113/9744 & $24.4\times13.3, -62.6$  & $0.14$  \\
08-07-2023\footnotemark[2]    & 0:30/0:09  & 113/9237 & $15.6\times14.3, 57.4$  & $0.19$  \\
18-07-2023    & 0:39/0:11 & 230/15238 & $15.8\times8.8, 39.6$  & $0.38$  \\
27-07-2023    & 0:36/0:10  & 230/16196 & $12.2\times9.1, -23.4$  & $0.51$  \\
02-08-2023    & 1:41/0:36  & 230/16196 & $12.3\times8.3, -29.7$  & $0.31$  \\
\hline
17-08-2023\footnotemark[3] & 00:59/00:22 & 83/14852 & $18.7\times14.6, 34.4$ & $0.05$ \\
\botrule
\end{tabular}
\footnotetext[1]{Superuniform weighting}
\footnotetext[2]{Data marked as {\it semi-pass} in ALMA data quality assessment}
\footnotetext[3]{Observations in ALMA Band 6}
\end{table}

\begin{figure}[ht]
\centering
\includegraphics[width=0.8\textwidth]{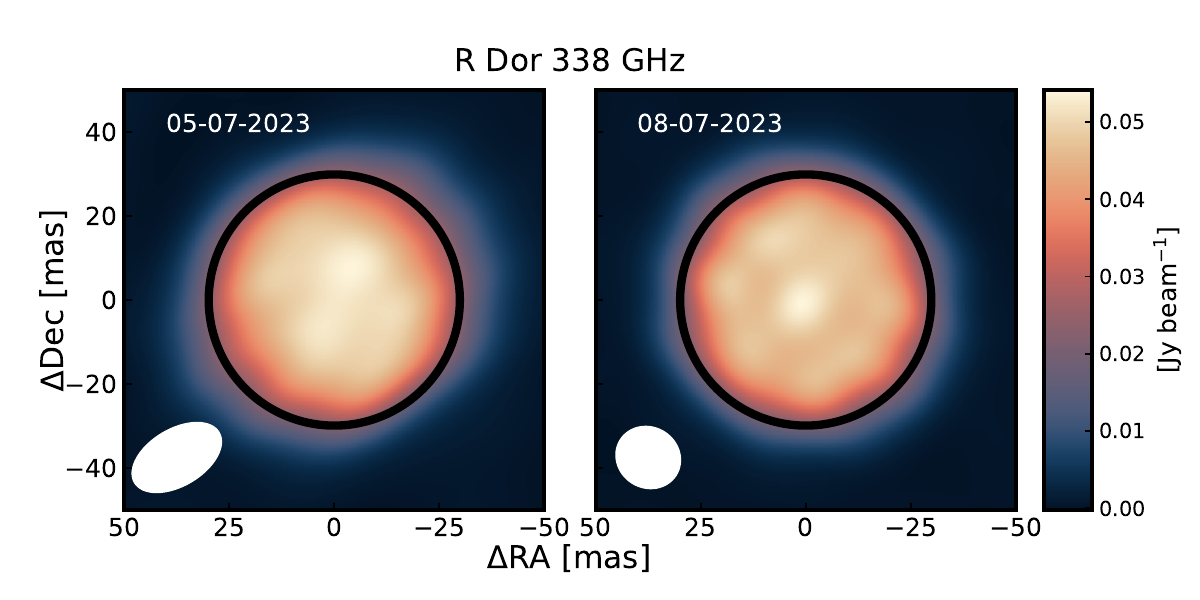}
\caption{The stellar surface of the AGB star R~Doradus. The panels represent the first two epochs of ALMA observations at 338~GHz. The black ellipse in the panels indicates the average size of the stellar disc at this frequency. The size and orientation of the interferometric beam is indicated at the bottom left of each panel.}\label{fig: firstepochs}
\end{figure}

\begin{figure}[h]
\centering
\includegraphics[width=0.8\textwidth]{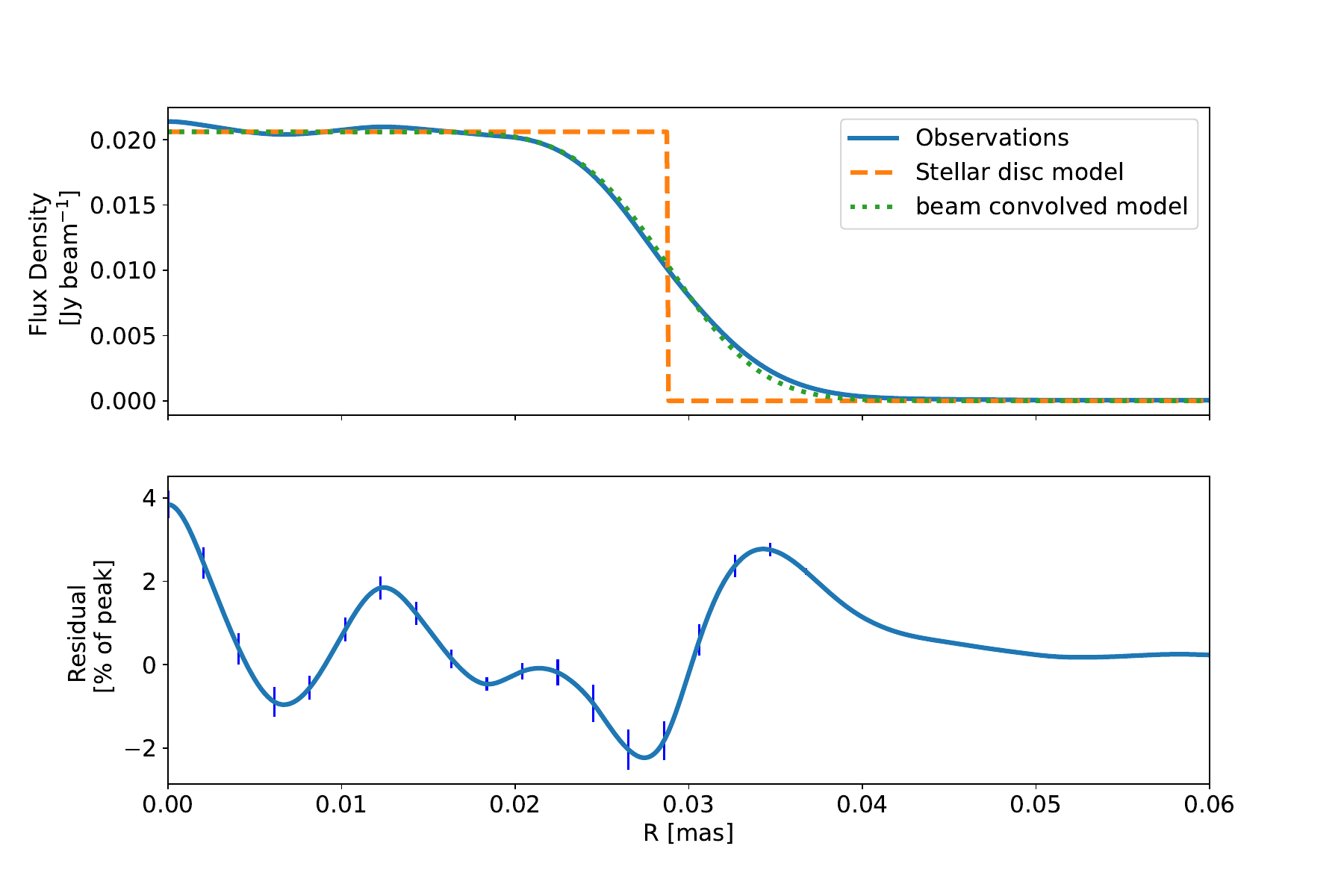}
\caption{The observed azimuthally averaged radial profile of the stellar disc of R~Doradus. The radial profile, in the top panel, is generated from a combination of the three last epochs. The dashed line indicates the uniform disc model that minimizes the residual flux shown in the bottom panel. The error bars on the radial profile are close to the line width, with a largest error of $5\times10^{-5}$~Jy~beam$^{-1}$. The error bars on the residual fraction are indicated with the vertical line segments.}\label{fig: disc}
\end{figure}

\begin{figure}[ht]
\centering
\includegraphics[width=0.8\textwidth]{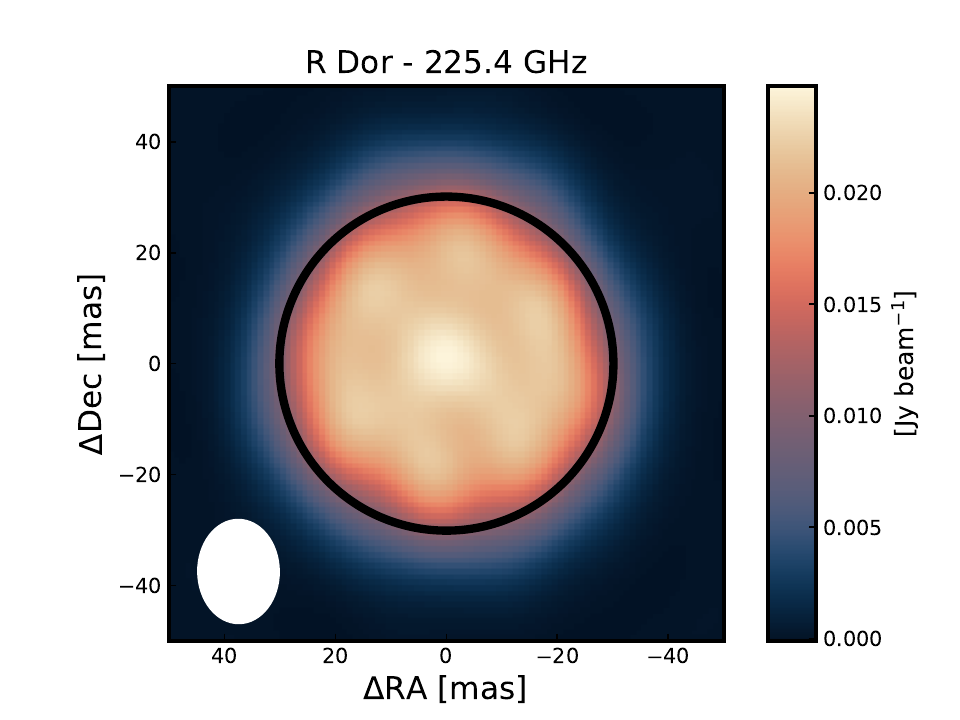}
\caption{The stellar surface of R~Doradus at 225~GHz. The observations
  were taken using ALMA band 6 on August 17 2023. The black ellipse in
  the panels indicates the average size of the stellar disc at
  225~GHz.}\label{fig: band6}
\end{figure}

\begin{figure}[t]
\centering
\includegraphics[width=0.8\textwidth]{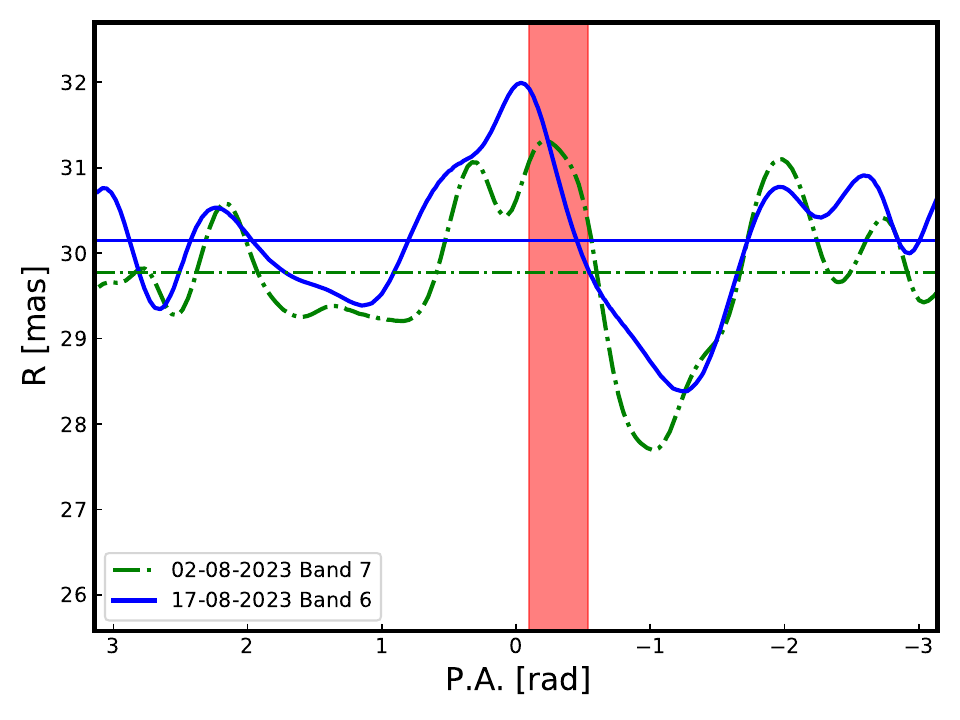}
\caption{As Fig.~\ref{fig: radvel}a, the half-power radius of R~Doradus as a function of
  the position angle. The curves indicate the half-power
  radius measured for the final epochs of the ALMA Band 7 (338~GHz)
  observations compared with the ALMA Band 6 (225~GHz) observations taken 15 days later. The red bar is the granule size derived from the higher
  resolution Band 7 observations. The improved signal-to-noise in Band 6 compensates for the lower flux and larger beam so that the typical $1\sigma$ s.d. radius uncertainty is also of order $0.2$~mas.}\label{fig: band6R}
\end{figure}

\clearpage





\end{document}